\documentclass{aa}  
\usepackage{xcolor}
\usepackage{graphicx}
\usepackage{txfonts}
\usepackage{newtxtext,newtxmath}

\usepackage[T1]{fontenc}
\usepackage{ae,aecompl}
\usepackage{natbib}

\usepackage{amsmath}	
\usepackage[normalem]{ulem}
\useunder{\uline}{\ul}{}
\usepackage{placeins}
\usepackage{rotating}
\usepackage{float}
\usepackage{threeparttable}
\usepackage[rightcaption]{sidecap}

\newcommand{\ks}{{$\rm K_s$}\ }

\begin{document}

   \title{Comparing bulge RR Lyrae stars with bulge giants}
   \subtitle{Insight from 3D kinematics}

   \author{J. Olivares Carvajal
          \inst{1,2}
          \and
          M. Zoccali\inst{1,2}
          \and
          M. De Leo\inst{2,1}
          \and
          R. Contreras Ramos\inst{1,2}
          \and
          C. Quezada\inst{1,2}
          \and
          Á. Rojas-Arriagada\inst{3,2,4,5}
          \and
          E. Valenti\inst{6,7}
          \and
          R. Albarrac\'in\inst{1,2}
          \and
          Á. Valenzuela Navarro\inst{1,2}
          }

   \institute{Instituto de Astrof\'isica, Pontificia Universidad Cat\'olica de Chile, Av. Vicu\~na Mackenna 4860, 782-0436 Macul, Santiago, Chile\\
              \email{jrolivares@uc.cl}
         \and
             Millennium Institute of Astrophysics, Av. Vicu\~na Mackenna 4860, 82-0436 Macul, Santiago, Chile
         \and
             Departamento de F\'isica, Universidad de Santiago de Chile, Av. Victor Jara 3659, Santiago, Chile
         \and
             N\'ucleo Milenio ERIS
         \and
             Center for Interdisciplinary Research in Astrophysics and Space Exploration (CIRAS), Universidad de Santiago de Chile, Santiago, Chile
         \and
             European Southern Observatory, Karl Schwarzschild-Strabe 2, D-85748 Garching bei Munchen, Germany 
         \and
             Excellence Cluster ORIGINS, Boltzmann-Strasse 2, D-85748 Garching Bei Munchen, Germany
             }

   \date{Accepted XXX. Received YYY; in original form ZZZ}

 
  \abstract
   {The structure and kinematics of the old component of the Galactic bulge are still a matter of debate. It
is clear that the bulk of the bulge as traced by red clump stars includes two main components, which are usually 
identified as the metal-rich and metal-poor components. They have different shapes, kinematics, mean metallicities, and alpha-element abundances. It is our current understanding that they are associated
with a bar and a spheroid, respectively. On the other hand, RR Lyrae variables trace the oldest population 
of the bulge. While it would be natural to think that they follow the structure and kinematics of the
metal-poor component, the data analysed in the literature show conflicting results.}
   {We aim to derive a rotation curve for bulge RR Lyrae stars in order to determine that the old component traced by these stars is distinct from the two main components observed in the Galactic bulge.  }
   {This paper combines APOGEE-2S spectra with OGLE-IV light curves, near-infrared photometry, and proper motions from 
the VISTA Variables in the V\'\i a L\'actea survey for 4197 RR Lyrae stars. Six-dimensional phase-space coordinates were used to calculate orbits within an updated Galactic potential and to isolate the stars.}
   {The stars
that stay confined within the bulge represent 57$\%$ of our sample.
Our results show that bulge RR Lyrae variables rotate more slowly than metal-rich red clump stars and have a lower velocity
dispersion. Their kinematics is compatible with them being the low-metallicity tail of the metal-poor component. We confirm that a rather large fraction of halo and thick disc RR Lyrae stars pass by
the bulge within their orbits, increasing the velocity dispersion. A proper orbital analysis is therefore critical to isolate bona fide bulge
variables. Finally, bulge RR Lyrae seem to trace a spheroidal component, although the current data do now allow us to reach a firm conclusion about the spatial distribution. }
    {}

   \keywords{Galaxy: bulge -- Galaxy: formation -- Galaxy: kinematics and dynamics -- Galaxy: structure -- stars: variables: RR Lyrae -- proper motions
               }

   \maketitle
%

\section{Introduction}

In the context of the Milky Way (MW) formation, the bulge is of special interest as a dominantly old component including at least 25\%  of the total stellar mass within a highly concentrated region \citep[e.g.,][]{Cao2013,Valenti2016,portail2017, Simion2017}. Most studies of the bulge structure, kinematics, and chemical content are based on red clump (RC) stars since these are fairly good distance indicators because their magnitude depends only mildly on age. Therefore, they are used to isolate a sample largely dominated by bulge stars, as opposed to disc foreground. Additionally, the mean magnitude of RC stars correlates with the mean distance of bulge stars towards a given direction, yielding information on the 3D structure of the bulge. However, the same low sensitivity to age prevents us from considering which population might have formed before the other. 
In this sense, RR Lyrae (RRL) variables are beneficial and complementary. They yield precise distances to individual stars, and most importantly, they are only present in stellar populations older than 10 Gyr. They are commonly used as tracers of a clean old population, without contamination from populations in the 5-9 Gyr age range that might (or might not) be mixed in the RC region of the colour-magnitude diagram (CMD).

From observations, we know that the bulge RC stars trace a bar with a boxy/peanut (X-shape) structure \citep{McWilliamZocalli,Nataf2010,Saito2011,Cao2013,wegg2013,Gonzalez2015,Ness2016}. This structure can be well explained by an in situ scenario considering a secular evolution of the disc \citep{DiMatteo2016} and the buckling of a bar \citep{CombesSanders1981} together with a kinematic fractionation \citep{Debattista2017}. The latter explains that in initial conditions, a radially cooler population (i.e. low-velocity dispersion) produces a peanut structure in the bulge over an extended period. Conversely, a radially hotter population (i.e. high-velocity dispersion) produces a boxy/spherical distribution. Observations from \cite{Zoccali2017, Zoccali2018}, and \cite{Lim2021} highlighted the presence of a spheroidal component in the bulge related to its metal-poor population.

Regarding kinematics, prior research has demonstrated that bulge RC stars exhibit cylindrical rotation \citep{Howard2009,Kunder2012,Ness2013,Zoccali2014A} displaying faster rotation for metal-rich stars than for metal-poor stars. Surprisingly, similar findings were observed for bulge main-sequence (MS) stars by  \cite{Clarkson2018}. Using Hubble Space Telescope (HST) photometry, these authors differentiated between metal-poor and metal-rich bulge MS stars. Combining these data with HST proper motions (PMs), they derived the transverse velocities of bulge MS stars in the Sagittarius Window Eclipsing Extrasolar Planet Search (SWEEPS) field, constructing a rotation curve that clearly indicated a distinct rotational behaviour between the two primary populations. Specifically, the metal-rich stars displayed substantial rotation compared to their metal-poor counterparts (see Fig. 11 in that study). 

Recent investigations focusing on very metal-poor stars in the bulge ([Fe/H]$<$$-$1 dex) have produced results congruent with those of the RC stars. The Pristine Inner Galaxy Survey (PIGS) first-data release \citep{Arentsen2020} offered insights into the kinematics of around 600 metal-poor stars in the bulge region, revealing a noticeable rotation gradient (and velocity dispersion) with metallicity: most metal-poor stars ([Fe/H]$<$$-$2 dex) showed negligible signs of rotation, but high-velocity dispersion. Several factors might account for this outcome. On the one hand, a pressure-supported component, such as a classical bulge, may dominate at the lowest metallicities \citep[e.g.][]{athanassoula+17}. On the other hand, contamination by halo interlopers must be taken into consideration for the more metal-poor stars.

Additionally, the Chemical Origins of Metal-poor Bulge Stars 2 (COMBS-2) survey \citep{Lucey2021} studied nearly 500 metal-poor stars in the bulge region. Using radial velocities (RVs) and PMs from Gaia Data Release 2 (DR2) \citep{Gaia2018}, the authors found that roughly 50\% of their sample were halo interlopers. After removing these stars, bulge metal-poor stars showed a slower rotation and a lower velocity dispersion than halo and bulge metal-rich RC stars. Therefore, the halo interlopers are crucial in increasing the velocity dispersion. The problem with the bulge metal-poor stars studies is mainly the need for more precise distance determinations in order to select the bona fide bulge-confined stars. 

From the observations, RRL stars are pulsating variable stars with periods between 0.2 and 1 day. They follow a period-luminosity-metallicity (PLZ) relation in the near-infrared (near-IR), which means that we can use them as standard candles to obtain precise distance measurements (uncertainties <10\%). They trace an old and metal-poor population since they are in the instability strip (IS) of the horizontal branch (HB). Although they represent only 1\% of the Galactic stellar mass \citep[e.g.,][]{Pietrukowicz2012,Nataf2013}, bulge RRL stars are considered the ancient stars in the MW Galaxy, 
meaning that the study of their kinematics and structure provides insights into the first stages of the formation of the Galaxy.  

The BRAVA-RRL survey \citep{Kunder2016} was the first survey dedicated to the kinematics of bulge RRLs. Using RVs for $\sim 1000$ RRLs, they found out that bulge RRL stars do not rotate and show a high velocity dispersion, probably due to a spheroidal component (pressure supported). More recently, \cite{Du2020} studied more than 15000 RRL stars by combining photometry from OGLE-IV with PMs from Gaia DR2, for which they derived tangential velocities and distances and obtained rotation curves for this old component. The rotation of the bulge RRLs was slower than in RC stars. The authors also showed that the metal-rich RRLs ([Fe/H]$>$$-$0.5 dex) rotate similarly to the bar and have a similar velocity dispersion as the metal-poor stars ([Fe/H]$<$$-$1 dex), which rotate more slowly. The Kunder et al. (2016) RRL sample lacks PMs and the Du et al. (2020) RRL sample lacks RVs, and therefore, neither survey was able to clean for halo stars. 
  
\cite{Prudil2019} were the first to obtain orbits for RRLs. With a sample of more than 400 RRLs and using both BRAVA-RRL RVs and Gaia DR2 PMs, they obtained orbital parameters for studying the kinematics of these stars. They found no evidence of rotation, even when they separated the two Oosterhoff groups \citep[related to metallicity,][]{Oosterhoff1939}. This confirmed the previous results that the old component rotates more slowly. Additionally, they found a halo contamination of 8 \%.
  
In \cite{Kunder2020} (hereinafter K20), the RRL sample was increased to $\sim 3000$ stars with RVs. Moreover, they complemented their data with Gaia DR2 PMs for $\sim 1400$ stars, performing an orbital analysis to find those confined to the bulge. They found that 25\% are halo interlopers, which is at odds with previous results. Considering an apocentre radius of 1.8 kpc, they separated the confined RRLs into inner (outer) and central RRLs to the Galactic centre. The result was that inner RRLs rotate and follow the bar behaviour, in contrast to central RRLs, which do not rotate. This result was confirmed in the review by \cite{Kunder2022} (hereinafter K22), where the sample with orbital parameters was increased to $\sim 2600$ stars. Moreover, for the central RRLs, the velocity dispersion sharply decreased after the halo decontamination, showing lower values than any other component and the slowest rotation. This is counter-intuitive to previous results. These results significantly affect the Galaxy formation models since even kinematic fractionation cannot explain how an initially cooler component lacks rotation in the end. 

Summarising, more information about the oldest stars in the Galactic bulge is necessary to confirm the latest observations and test the current formation models. The connection between RRL stars and the more frequently studied RC/RGB stars from the bulge is still unexplored. Thus, it is essential to work with larger statistics and obtain orbital parameters to reveal the actual behaviour of this old population.  

In this work, we study the kinematics of a bulge RRL sample by performing a 6D analysis, for which we mixed spectroscopic and photometric surveys in order to reveal the properties of the ancient population of the bulge. Section 2 describes the data provided from different surveys. Section 3 shows the analysis with which we estimated the distances. Section 4 explains the orbital analysis. In Section 5, we present the rotation curves. Section 6 shows the 3D spatial distribution of the stars. Finally, in Section 7, we present the discussion and conclusions.  
\begin{figure*}
	\includegraphics[width=15cm]{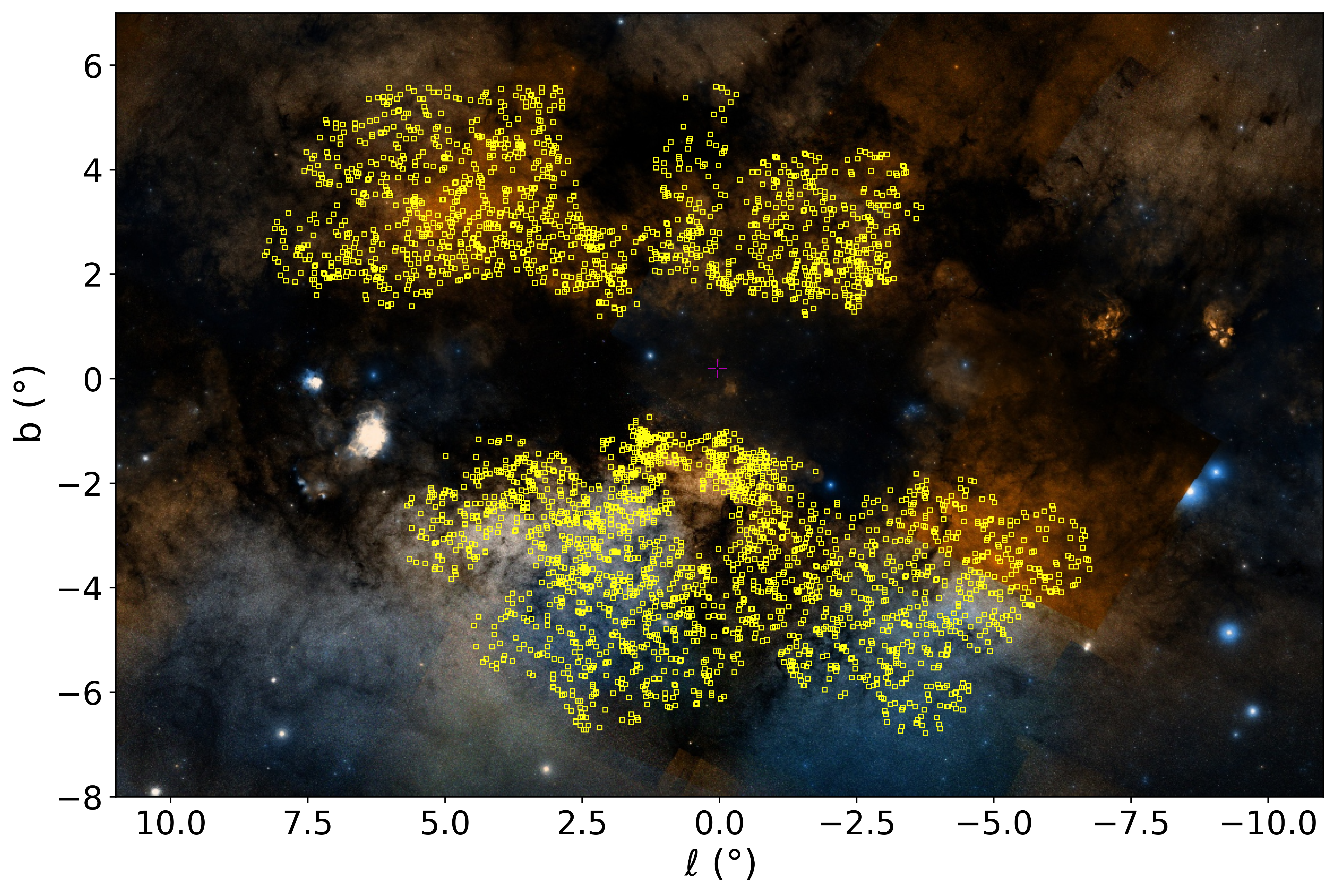}
    \caption{Position of the 4455 APOGEE-2S RRab variables, which have RVs, in Galactic coordinates. The targets (yellow) are overlaid on an image from Aladin using a red SDSS-2 map.}
    \label{fig:location}
\end{figure*}

\section{Data}
\label{sec:data}

\subsection{APOGEE-2S spectra}

Our initial sample of RRL stars came from the APOGEE Data Release 17 (DR17). The Apache Point Observatory Galactic
Evolution Experiment 2 South \citep[APOGEE-2S;][]{APOGEE2s} is a near-IR spectroscopic survey carried 
on at the Dupont 2.5m telescope, located at Las Campanas Observatory (LCO) in Chile. It has a spectral resolution of
R$\sim$22000 and a spectral range between 1.51 and 1.7 microns. We used data from the 17th 
data release \citep[DR17;][]{apogeeDR17}, selected by their RRLYRAE tag and by their location towards the Galactic bulge, as shown in Fig.~\ref{fig:location}. This selection yielded 5626 targets, which were originally selected as
confirmed RRLs from the OGLE-IV catalogue by \cite{ogleIV} (see \citet{zasowski+17} for details about the target selection).
Through the use of near-IR data, this is the first spectroscopic study of RRL in the northern Galactic hemisphere that is more affected by extinction.

\begin{figure}
	\includegraphics[width=\linewidth]{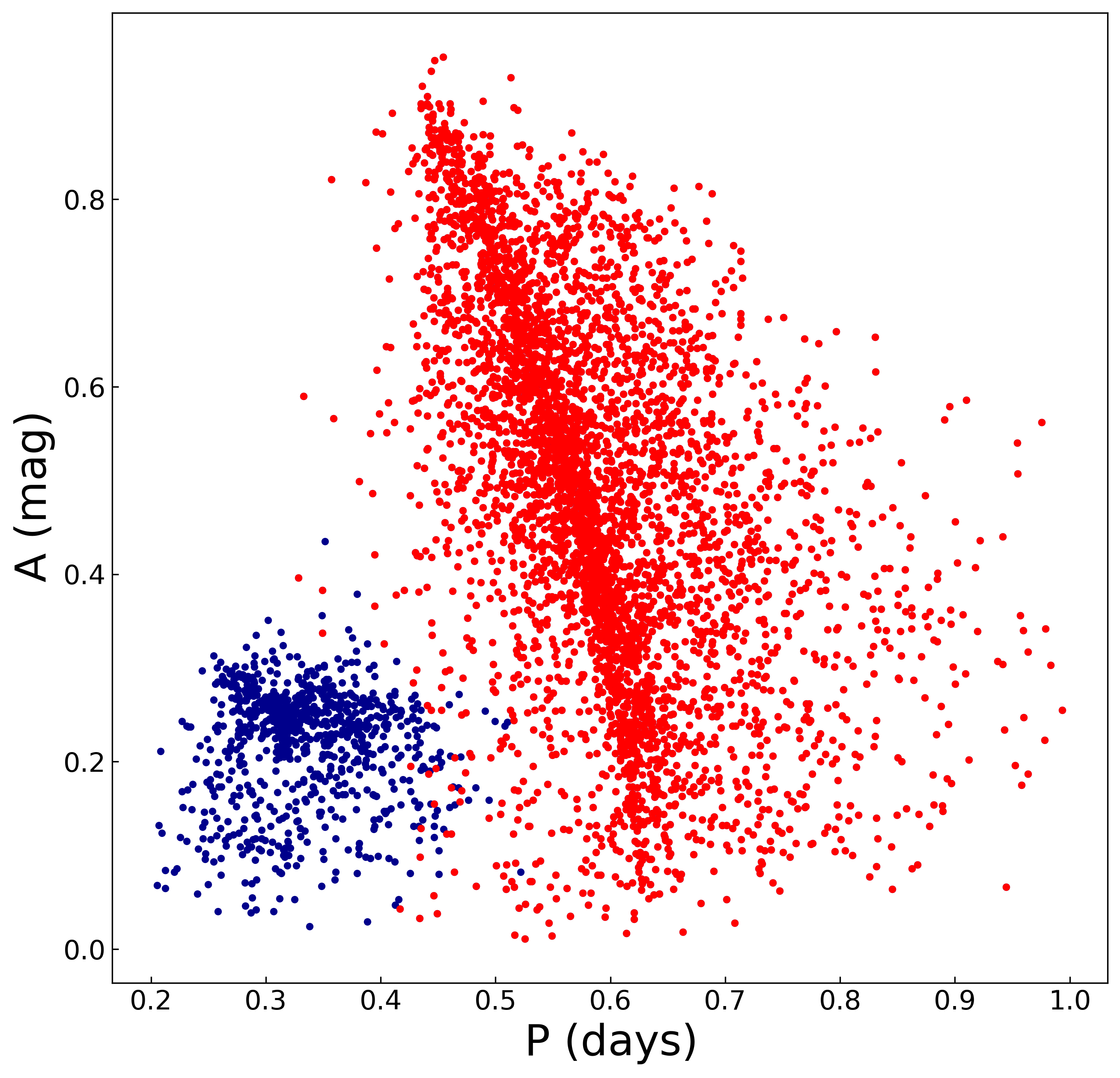}
    \caption{Bailey diagram for the APOGEE RRL stars using the period and amplitude in the I band from OGLE-IV. The RRab stars are plotted in red, and the RRc stars are shown in blue.}
    \label{fig:bailey}
\end{figure}

A few selections were imposed on the initial sample of 5626 RRL.
Stars with a low signal-to-noise ratio (S/N<5) were excluded, leaving us with 5351 RRLs. Additionally, we performed a cross-match with the OGLE-IV catalogue in order to select only 
fundamental mode (ab-type; RRab) variables as suitable to derive systemic RVs \citep{sesar+10}. For this step, we used the TYPE column in the OGLE-IV catalogue, retaining 4455 RRab stars. They are shown in red in Fig.~\ref{fig:bailey} (the Bailey diagram) together with the 896 discarded RRc, which are shown in blue.

\begin{figure*}
	\includegraphics[width=18cm]{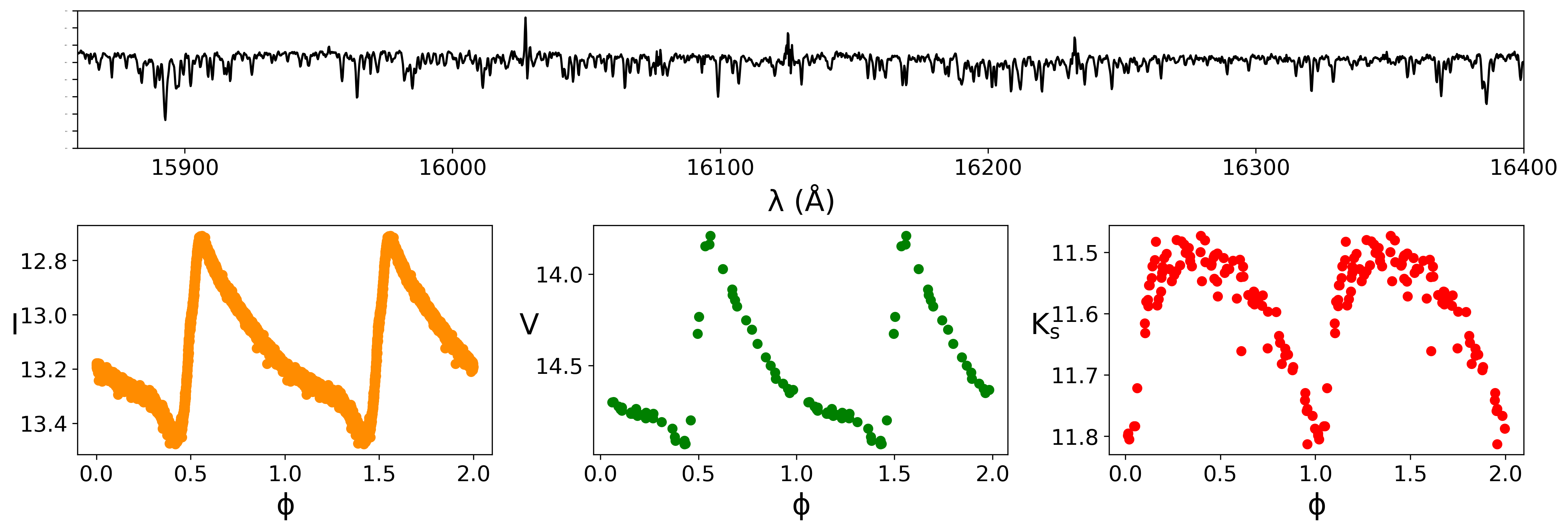}
    \caption{Example of the data available for the RRL star OGLE-BLG-RRLYR-32068 with a period P=0.4627031 days. \textit{Top panel:} APOGEE spectrum for the RRL in the H-band regime. \textit{Bottom panels:} From left to right, the light curves in the I, V (OGLE-IV), and $K_s$ (VVV) bands for the same RRL star.  }
    \label{fig:spec}
\end{figure*}
\subsection{Light curves}
\label{sec:LCs}

Both I-band and V-band light curves are available from OGLE-IV for each of our targets (Fig.~\ref{fig:spec}, bottom left and middle panels). 
The relevant parameters of the I-band light curve, such as period and amplitude, are included in the OGLE RRLyr catalogue. The V-band amplitudes, in contrast, are not included in the catalogue, but are needed for the calculation of the systemic RVs. Therefore, we downloaded the V-band light curves available online, fitted them with the templates provided by \citet{sesar+10}, and also derived Fourier parameters for each variable. 

The VISTA Variables in the V\'\i a L\'actea survey \citep[VVV][]{minniti10} used the VISTA 4.1m telescope at the European Southern Observatory (ESO) in Paranal, equipped with the VIRCAM near-IR camera to observe the Galactic bulge and disc for more than nine years. We extracted the J-, H-, and \ks-band light curves for most of our targets from the point spread function (PSF) photometry derived as described in \citet{Contreras2018}. While the \ks light curves contain $\sim$80 epochs (Fig.~\ref{fig:spec}, bottom right), the VVV survey in the J and H bands only includes a few epochs, so that their light curves have been omitted in the figure. 

    \begin{figure*}
        \includegraphics[width=18cm]{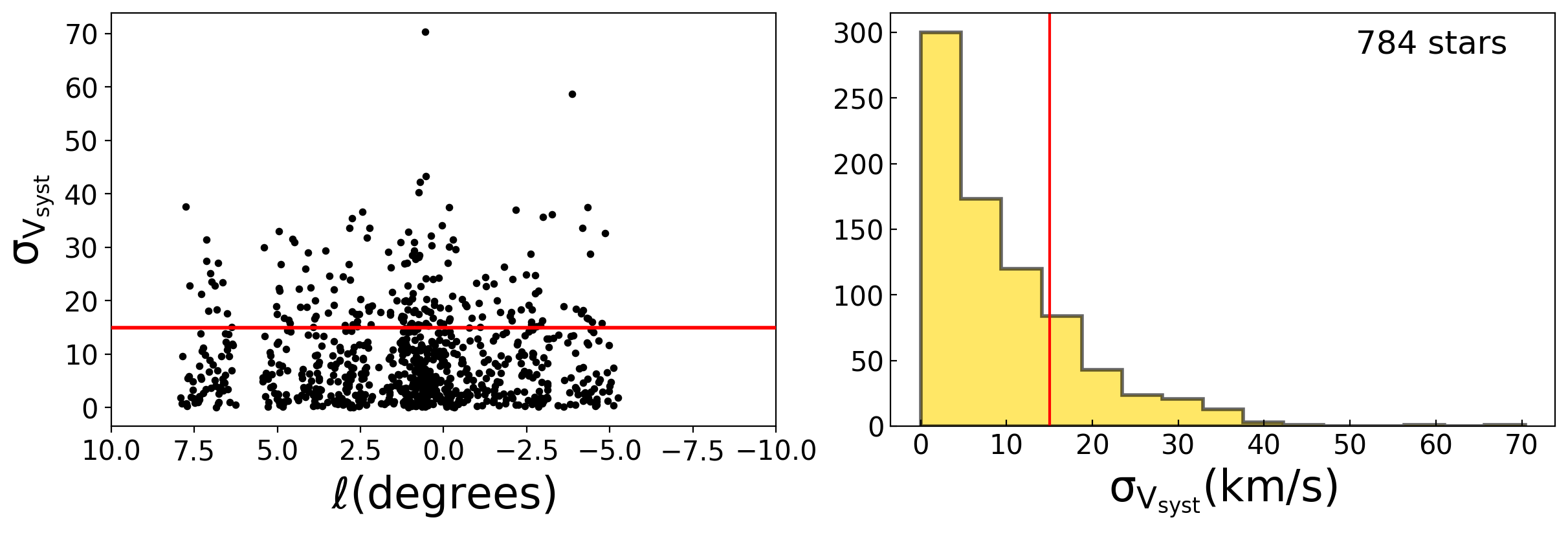}
         \caption{Diagnostic plots of the obtained systemic RVs. \textit{Left:} Dispersion of multiple estimates of the systemic RVs for 784 stars with more than one visit along the Galactic longitude coordinate. The red line shows the typical velocity error of $\rm 15 \ km \ s^{-1}$ from \cite{sesar+12}. There is no trend present in the space. \textit{Right:} Histogram with the dispersion distribution.}
        \label{fig:rvcheck}
    \end{figure*}
\subsection{Systemic radial velocities}
\label{sec:RVs}

The APOGEE-2S survey observed several stars more than once and up to three times (visits). The RV of each visit is available in the APOGEE DR17 catalogue. A median value from multiple visits is also available, but is not useful for pulsating variable stars since these stars change their velocity due to pulsation. Instead, a systemic RV, or barycentric RV ($\rm RV_{\gamma}$), needs to be derived that takes the pulsation velocity at a given phase into account and removes it. To this end, we used 
RV curves and the equation from \citet{sesar+12}, which was based on an extensive study of six RRab variables,
\begin{equation}
\rm    RV_{\gamma}=RV_{obs}(\Phi_{obs})-A_{RV}\times T(\Phi_{obs}),
\end{equation}
where RV$_{\rm obs}$ is the observed RV at phase $\Phi_{\rm obs}$, T($\Phi_{\rm obs})$ is the value in the normalized RV curve template of an RRab star at that phase, and A$_{\rm RV}$ is the amplitude of that RV curve. In the same study, $\rm A_{RV}^{met}$ is defined as
\begin{equation}
\rm    A_{RV}^{met}=25.6(\pm 2.5)\times A_V+35.0(\pm2.3), ~~~~~~~ \sigma_{fit}=2.4 \ km \ s^{-1},
\end{equation}
where A$_{\rm RV}^{\rm met}$ is the amplitude of the RV curve based on metallic optical lines from the same work, and A$_{\rm V}$ is the amplitude of the light curve in the V band derived above. These relations allowed us to derive a systemic RV even for the variables with only one visit as long as the observation phase is known.

We have two visits for 734 stars and three visits for 50 stars. We used them to test the quality of the systemic RV estimates. In order to obtain something more quantitatively robust, we compared the dispersion of the different measurements of the systemic RV using eq. (1) versus longitude. Fig. \ref{fig:rvcheck} shows the result, where we did not find evidence of a trend with longitude. The red line represents the error of $\rm 15 \ km \ s^{-1}$ estimated by \cite{sesar+12}. Furthermore, the right panel of Fig. \ref{fig:rvcheck} shows the dispersion distribution of the stars, where it is evident that the majority of the data lies within $\pm 1\sigma$. Moreover, the two outliers in the left panel can be explained as outside $\pm 3\sigma$ of the distribution, and we maintained them to preserve the original statistics. Based on these data, we conclude that this sub-sample is a suitable representation of the complete sample. We defined the error in the systemic RV as the squared sum of the error quoted by \citet{sesar+12} ($\pm \rm 15 \ km \ s^{-1}$) and the observed RV error for each star.

We also compared our results with the RV estimates from the work of K20, who obtained the observed RVs using the AAOmega multi-fibre spectrograph in the Anglo-Australian 3.9m Telescope with a resolution of $R \sim 10000$. We had 587 stars in common, and we have similar RV estimates for most cases. We found a systematic of $\rm 12 \ km \ s^{-1}$ compared with the K20 systemic RVs. K20 defined the systemic RV as the one at phase $\Phi=0.38$; on the other hand, \cite{sesar+12} defined the systemic RV as the one at phase $\Phi=0.5$. Therefore, this difference in the definitions of the systemic RVs produces the observed offset. We also derived a dispersion of $\rm \sigma = 24 \ km \ s^{-1}$ for the difference in the measured systemic RVs. This value exceeds the error of \cite{sesar+12}; however, we used a different approach to that employed in K20 to obtain the systemic RVs. 

Nonetheless, the RV of 59 stars ($\sim$ 10\%) differs by more than 80 $\rm km \ s^{-1}$. This difference is significantly larger than the expected difference because the observations were taken at different pulsation phases, and therefore, it deserves a closer inspection. It turns out that 16 stars (27\% of the outliers) are tagged with a flag over 0 in K20, meaning that their estimate of the RV was considered poor. For the other stars, 6 (10\%) have a constant RV at different epochs in K20, however, suggesting that the light in the fibre could be dominated by a non-pulsating star. We note that the fibre diameter in the AAOmega spectrograph is 2 arcsec, while 
it is 1.3 arcsec in APOGEE-2S. Therefore, the possibility of significant contamination by another star is higher in the former. Finally, a third group of 38 stars (64\%) has only one visit in APOGEE and multiple visits in K20, suggesting that the K20 velocity could be more reliable. Because these are a small fraction of our total sample and we lack a clear indication of a mistake in our data, we did not remove them, but rather checked that their position was evenly distributed in longitude. Thus, they do not introduce a bias in the derived rotation curves.

    \begin{figure*}
        \includegraphics[width=18cm]{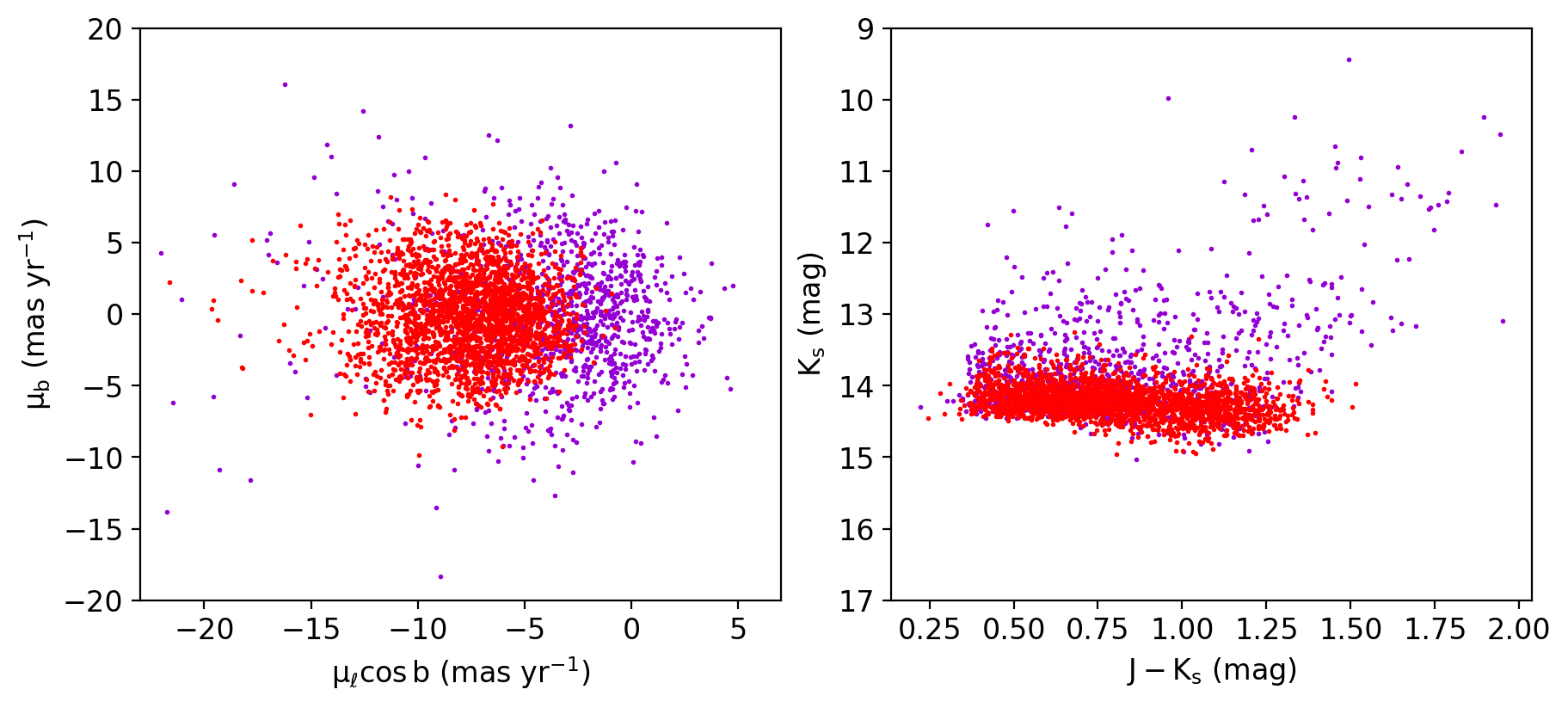}
         \caption{VVV photometry and PMs for the bulge RRLs. \textit{Left:} VPD in Galactic coordinates for the stars of our sample. The red points represent the bulge-confined RRLs, and the purple points show the interlopers. \textit{Right:} CMD of the sample using the same symbols. It is evident that the bulge-confined RRLs are less dispersed around the HB in the CMD, proof that our selection criteria are appropriate.}
        \label{fig:VPDCMD}
    \end{figure*}

\subsection{Proper motions}
\label{sec:PMs}

Near-IR PSF photometry and PMs for our sample of RRL stars were acquired from the VVV Survey using the approach outlined by \cite{contrerasramos17}. Furthermore, the PMs were meticulously calibrated to the Gaia astrometric system by means of all the common stars within each VVV detector, ensuring the determination of accurate absolute PMs.

When searching the VVV catalogue, we identified 4331 RRLs, with 124 stars outside the VVV field of view (FoV). PMs were successfully derived for 4197 stars because certain stars were subject to saturation since the VISTA Telescope (4.1m) is significantly larger than the Warsaw University Telescope (1.3m) that is used by OGLE.

Fig. ~\ref{fig:VPDCMD} depicts the vector point diagram (VPD), where the sample is clustered around the proper motion (PM) values of the Galactic bulge $(\mu_\ell \cos{b}, \mu_b)=(-6,0) \ \rm{mas \ yr^{-1}}$. The CMD of the RRL sample is also presented in the figure mentioned above. This CMD is based on the mean magnitudes in the Ks and J bands rather than on a singular catalogue value. The method employed to calculate these mean magnitudes is detailed in the subsequent section. All stars are clearly situated within the HB region, as expected for RRL stars undergoing helium burning within the IS. The PMs and their errors are included in Table \ref{tab:table1}.

\begin{table*}
\centering
\scriptsize
\caption{Observational parameter of the RRLs in our sample.\protect\footnotetext{Tables 1 and 2 are only available in electronic form at the CDS via anonymous ftp to cdsarc.u-strasbg.fr (130.79.128.5) or via http://cdsweb.u-strasbg.fr/cgi-bin/qcat?J/A+A/.}}
\begin{tabular}{llllllllllllll}
\hline
OGLE ID & $\alpha$   & $\delta$   & $P$       & $V_\gamma$     & $V_{GC}$       & $\langle K_s \rangle$ & $\langle J \rangle$ & {[}Fe/H{]} & e{[}Fe/H{]} & $\mu_\ell \cos{b}$    & $\mu_b$               & D     & Flag \\
        & (deg)      & (deg)      & (days)    & $(km\ s^{-1})$ & $(km\ s^{-1})$ & (mag)                 & (mag)               & (dex)      & (dex)       & ($\rm mas \ yr^{-1}$) & ($\rm mas \ yr^{-1}$) & (kpc) &      \\ \hline
00369   & 260.889583 & -29.289778 & 0.4681714 & -57            & -59            & 14.246                & 14.903              & -1.36      & 0.01        & -3.426                & -10.666               & 7.3   & 0    \\
00372   & 260.899333 & -29.370972 & 0.5135301 & -31            & -33            & 14.265                & 14.901              & -1.38      & 0.02        & -9.608                & -2.299                & 7.8   & 0    \\
00397   & 260.969292 & -29.279111 & 0.5369132 & -89            & -92            & 14.318                & 14.931              & -1.22      & 0.02        & -8.939                & -1.406                & 8.2   & 0    \\
00400   & 260.984417 & -29.213694 & 0.6708366 & -73            & -75            & 14.197                & 14.883              & -1.89      & 0.02        & -3.057                & -1.812                & 8.8   & 0    \\
00409   & 261.025833 & -29.298889 & 0.6232191 & 338            & 336            & 13.886                & 14.53               & -1.42      & 0.05        & 5.683                 & 2.618                 & 7.4   & 0    \\
00418   & 261.061542 & -29.250500 & 0.5643130 & -101           & -103           & 14.134                & 14.787              & -1.14      & 0.04        & -10.424               & -1.61                 & 7.8   & 0    \\
00427   & 261.082219 & -29.118687 & 0.6619568 & -212           & -214           & 13.677                & 14.356              & -2.08      & 0.02        & -3.402                & 5.677                 & 7.0   & 0    \\
00431   & 261.089000 & -29.203639 & 0.6082388 & 86             & 85             & 14.121                & 14.76               & -1.90      & 0.04        & -7.57                 & 1.888                 & 8.1   & 0    \\
00468   & 261.197083 & -29.576194 & 0.4633249 & 27             & 24             & 14.226                & 15.005              & -0.56      & 0.25        & -5.768                & -4.778                & 7.1   & 0    \\
00486   & 261.230849 & -29.380247 & 0.504192  & 233            & 231            & 12.468                & 13.704              & -1.35      & 0.45        & -7.704                & -0.074                & 3.0   & 1    \\ \hline
\end{tabular}
\begin{tablenotes}
\item All IDs start with the OGLE-BLG-RRLYR identifier. The coordinates, periods, systemic RVs, galactocentric RVs, mean magnitudes, metallicities, proper motions, and distances are presented. The complete version, which includes the errors, is only available in electronic form at the CDS via anonymous ftp to cdsarc.u-strasbg.fr (130.79.128.5) or via http://cdsweb.u-strasbg.fr/cgi-bin/qcat?J/A+A/.
\end{tablenotes}
\label{tab:table1}
\end{table*}

\section{Distances}
\label{sec:dist}

To determine distances using RRL stars, it is essential to have reliable absolute magnitudes, mean magnitudes, and reddening values. The following section details the analysis process to derive these parameters.

\subsection{Mean magnitudes}
We used VVV near-IR light curves to calculate the mean magnitudes in the J and $\rm K_s$ bands. In the $\rm K_s$ band, we have an average of 80 data points per light curve, and in the J-band averages, we have approximately 3 data points spread across 9 years. In the VVV eXtension region \citep[VVVX][]{Minniti2018}, these averages increase to 115 and 9 for the $\rm K_s$ and J bands, respectively, over a baseline of roughly 13 years.

Initially, we corrected the raw observations for the sample of 4192 stars using different factors. The light curves consist of two columns: The first column represents the modified Julian day (MJD), which must be corrected for Earth's motion, especially given the comparatively short periods of RRLs compared to other variable stars. We employed the Python package {\tt WCS} to convert MJD into heliocentric Julian day (HJD) for each light curve. 

The second column contains the apparent magnitudes in either the J or the $\rm K_s$ band. These magnitudes were calibrated to the 2MASS photometric system using the appropriate zero point \citep{GonzalezFernandez2018}.

To derive mean magnitudes, we employed two codes from the literature. The {\tt lcfit} code \citep{Dekany2019} for robust regression of periodic time series provides a $\rm K_s$ mean magnitude and an error. This code is only available for the $\rm K_s$ band; therefore, we used the {\tt PyFiNeR} code \citep{Hajdu2018}, which uses RRL near-IR light-curve fitting techniques to determine robust mean magnitudes for the J band (but without errors).

\begin{figure*}
	\includegraphics[width=\linewidth]{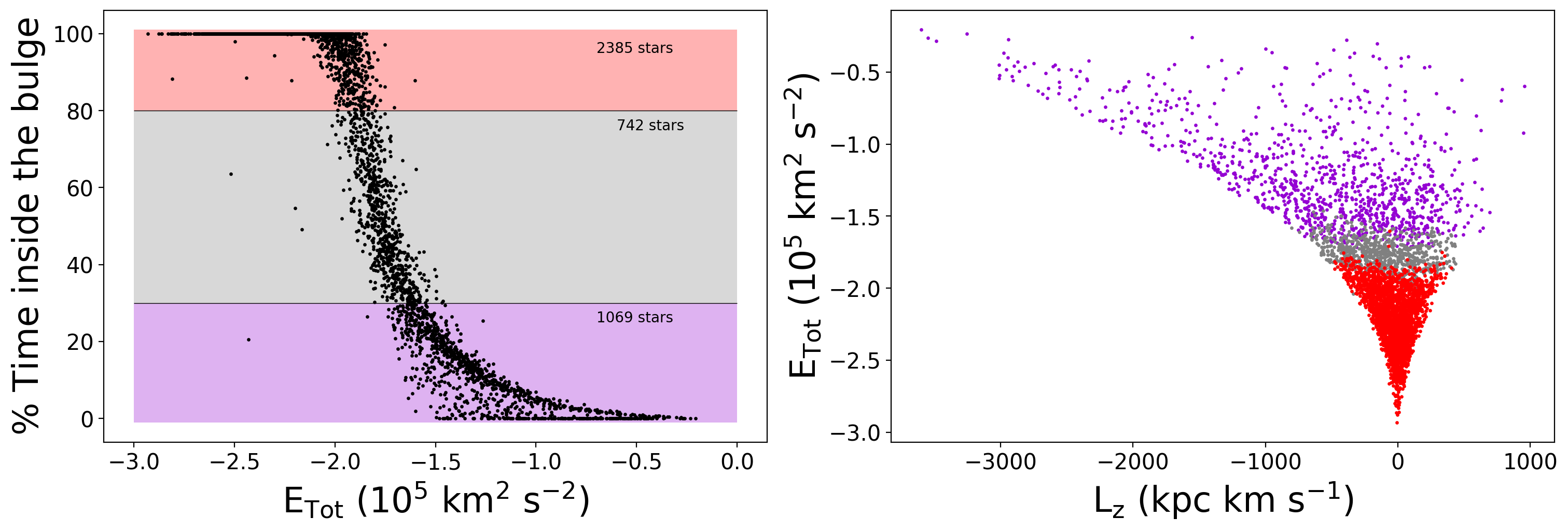}
    \caption{Orbital selection of the RRLs. \textit{Left panel}: Classification of the RRL sample considering the time inside the bulge radius. Confined bulge RRLs lie in the red area, RRLs with an unknown-origin lie in the grey area, and halo and disc RRLs lie in the purple area. \textit{Right panel}: Total energy vs the angular momentum of the RRL stars. The bulge-confined RRLs (red) have less energy than the halo and disc (purple) and unknown-origin (grey) interlopers.}
    \label{fig:orbselec}
\end{figure*}

\begin{table*}
\centering
\caption{Orbital parameters derived for the RRLs of the sample.}
\begin{tabular}{llllllll}
\hline
OGLE ID & E$\rm _{Tot}$           & L$_z$                     & $R_A$ & $R_P$ & $\rm Z_{max}$ & $e$  & $\rm T_{2.5 kpc}$ \\
        & ($\rm km^{2} \ s^{-2}$) & ($\rm kpc \ km \ s^{-1}$) & (kpc) & (kpc) & (kpc)         &      & (\%)              \\ \hline
00369   & -108202.1               & -369.9                    & 11.8  & 0.8   & 6.9           & 0.87 & 7                 \\
00372   & -237533.8               & -67.7                     & 0.8   & 0.2   & 0.7           & 0.59 & 99                \\
00397   & -242076.7               & -43.0                     & 0.8   & 0.3   & 0.6           & 0.39 & 99                \\
00400   & -183713.8               & 130.3                     & 3.3   & 0.6   & 2.7           & 0.67 & 47                \\
00418   & -236968.2               & -84.0                     & 0.8   & 0.5   & 0.5           & 0.22 & 99                \\
00427   & -120865.4               & -554.1                    & 11.5  & 2.0   & 4.5           & 0.70 & 4                 \\
00431   & -230991.8               & -4.5                      & 1.0   & 0.3   & 0.9           & 0.48 & 99                \\
00468   & -181826.7               & -315.2                    & 2.7   & 0.6   & 1.9           & 0.62 & 72                \\
00486   & -70462.66               & -1794.8                   & 20.4  & 2.8   & 4.9           & 0.76 & 0                 \\
00520   & -156809.7               & -346.4                    & 3.9   & 0.6   & 3.1           & 0.71 & 36                \\ \hline
\end{tabular}
\scriptsize
\begin{tablenotes}
\item All IDs start with the OGLE-BLG-RRLYR identifier. The total energy, angular momentum, apocenter radius, pericenter radius, maximum height, eccentricity, and time percentage within the bulge are presented. The complete version, which includes the errors, is only available in electronic form at the CDS via anonymous ftp to cdsarc.u-strasbg.fr (130.79.128.5) or via http://cdsweb.u-strasbg.fr/cgi-bin/qcat?J/A+A/.
\end{tablenotes}
\label{tab:table2}
\end{table*}
\subsection{Metallicities}
\label{sec:met}

We determined photometric metallicities using the {\tt rrl\_feh} code for estimating metallicity in RRL stars through deep learning \citep{Dekany2022}. This code employs Fourier fitting techniques to derive the photometric metallicity from the $K_s$-band light curve. We used all available $K_s$-band light curves for the derivation of metallicities.

The {\tt lcfit} code, mentioned in the previous subsection, also provides S/N values for each light curve. This parameter is crucial for deriving other parameters from the light curve, such as metallicity. We inspected the fits by eye and realized that the method becomes unsatisfactory when the light-curve S/N falls below 40. Consequently, we labelled stars with S/N<40 in the final catalogue. In the "Flag" column, a value of 1 denotes a good S/N, while 2 means an S/N below 40. The metallicities and the flags for each RRL star are listed in Table ~\ref{tab:table1}.

\subsection{Distance determination}

To determine the distances, we used the classic distance equation
\begin{equation}
    K_s-M_{K_s}-A_{K_s}= 5 \log{d}-5,
\end{equation}
where $K_s$ is the mean magnitude, $M_{K_s}$ is the absolute magnitude, and $A_{K_s}$ is the extinction coefficient. The absolute magnitude was calculated using the period-luminosity (PL) relation by \citet{Neeley2019} for the $K_s$ band. Zoccali et al. (2024) showed that in the Galactic bulge, several PLZ in the literature produce a non-negligible segregation of the RRL, either in metallicity or in the period. The PL by \citet{Neeley2019}, in contrast, has proven to minimize these effects, ensuring at least that relative distances would be self-consistent,  
\begin{equation}
   M_{K_s}=(-0.37\pm0.35)-(2.84\pm0.35)\times(\log{P_F}+0.3),
\end{equation}
where $P_F$ is the fundamental period of the RRab star. The extinction coefficient can be calculated as
\begin{equation}
    A_{K_s}=E(J-K_s)\times R_{K_s},
\end{equation}
where $R_{K_s}$ is the extinction ratio. We adopted $R_{K_s}=0.465$ from \cite{minniti2020}. The reddening was calculated as
\begin{equation}
    E(J-K_s)=(J-K_s)-(J-K_s)_0,
\end{equation}
where $(J-K_s)_0=0.26\pm0.03$ mag is the mean intrinsic colour for RRab stars from \cite{Contreras2018}, and $J$ and $K_s$ are the mean magnitudes calculated in Sect. 3.1. 

In order to estimate the statistical error on the distances, we used the following formula, derived from the distance modulus expression:
\begin{equation}
    (\Delta \mathrm{d})^2= (0.46 \mathrm{d})^2 [\delta K_{s}^2+\delta A_{K}^2 + \delta M_K^2],
\end{equation}
where $\delta K_{s}$ is the mean magnitude error in the $K_{s}$ band, $\delta A_{K}$ is the extinction error, and $\delta M_K$ is the absolute magnitude error. The value for $\delta K_{s}$ was obtained using the {\tt lcfit} code, while $\delta M_K$ was propagated from the PL relation, using  the errors associated with the \textit{a} and \textit{b} coefficients. Since our target selection was based on the OGLE-IV catalogue, restricted to a relatively low extinction area, the $\delta A_{K}$ factor has a marginal impact on the error budget.

The statistical error calculated above is an underestimation of the total error because different PL relations yield distances with a spread comparable to it. To address this point, we derived distances using five recent PL(Z) relations available in the literature \citep{Muraveva2018,Neeley2019,Bhardwaj2023,Zgirski2023,Prudil2023}, calculated the standard deviation of the distances for each star, and used this value as an estimate of the systematic error. The total error is the squared sum of the statistical error and the systematic error. The calculated distance and the error for each RRL star are listed in Table ~\ref{tab:table1}\footnote{Tables 1 and 2 are only available in electronic form at the CDS via anonymous ftp to cdsarc.u-strasbg.fr (130.79.128.5) or via http://cdsweb.u-strasbg.fr/cgi-bin/qcat?J/A+A/.}.

\section{Orbital parameters}
\label{sec:orbits}

An orbital parameter analysis is crucial when analysing stellar kinematics because some stars could be only temporarily transiting by the inner Galaxy along their orbits.

Rather than using a publicly available tool to derive the orbital parameters, we developed our own orbit integrator (De Leo, in prep.). This gave us the freedom to build the model for the gravitational potential of the MW from scratch and make it as close as possible to current observed properties. Special care was devoted to the modelling of inner structures and their rotation, as their effect on the orbits of stars within the inner few kiloparsec is non-negligible \citep{2018AstBu..73..162C, 2023Galax..11...26T}. Details of this potential will be provided in De Leo et al. (in prep.), but we briefly describe it here. The model of the MW gravitational potential employed in our orbit integrator is composed of a Navarro-Frenk-White (NFW) dark matter (DM) halo \citep{1996ApJ...462..563N}, two Miyamoto-Nagai (MN) stellar discs \citep{1975PASJ...27..533M}, two MN gaseous discs, a Long-Murali (LM) rotating bar \citep{1992ApJ...397...44L}, and a spherical bulge component modelled as a Hernquist profile \citep{1990ApJ...356..359H}. The parameters of the DM halo are the same as were employed in the \texttt{MWPotential2014} from \citet{2015ApJS..216...29B} with a mass $M_{DM} = 8.0 \times 10^{11} M_{\odot}$. The two stellar MN discs were set up to reproduce the observed thin and thick disc of the MW, their scale heights were fixed at $z_{thin} = 0.3$ kpc and $z_{thick} = 0.9$ kpc, and their masses and scale lengths were chosen to recover the local mass densities measured by \citet{2022MNRAS.513.4130L}, giving us $M_{thin} = 3.65 \times 10^{10} M_{\odot}$, $M_{thick} = 1.55 \times 10^{10} M_{\odot}$, $a_{thin} = 3.5$ kpc, and $a_{thick} = 2.0$ kpc. The two gaseous MN discs represent the HI and molecular gas discs present in the MW with their masses (for a total gas mass of $M_{gas} = 1.22 \times 10^{10} M_{\odot}$) and scale heights fixed to the same values as the gas discs in \citet{2017MNRAS.465...76M} and their scale lengths set to reproduce the local mass densities measured for the respective gases (i.e. \citealt{2015ApJ...814...13M}). The bulge of our model was composed of both a bar and a spheroidal component. The LM bar had a mass of $M_{bar} = 10^{10} M_{\odot}$, was geometrically similar to the single-bar model in \citet{2015MNRAS.450.4050W}, and rotated with a pattern speed $\Omega_p = 41.3\pm3$\,km\,s$^{-1}$\,kpc$^{-1}$ \citep{2019MNRAS.488.4552S}. The spheroidal component has a mass of $M_{sph} = 10^{10} M_{\odot}$ and a scale radius of $a_{sph} = 0.3$ kpc. The presence of a massive spheroid was inferred from observations of the MW bulge stellar populations \citep{Valenti2016, Zoccali2018, Lim2021, 2021A&A...656A.156Q} and the spheroid is analogous to the one seen in the slightly larger companion of the MW, the M31/Andromeda galaxy \citep{2017MNRAS.466.4279B, 2018MNRAS.481.3210B}.

In order to derive the orbital parameters, we converted the observed galactic coordinates ($\ell$.b), the RVs, the PMs, and the calculated distances into the Cartesian galactocentric frame with the Astropy modules \citep{astropy1,astropy2}. We used a distance of the Sun to the Galactic centre of 8.34 kpc \citep{Reid2014}. We also used this value since our distance to the GC by RR Lyrae is closer to this one (8.3 kpc). For each star, the orbits were evolved backwards in time for 10 Gyr inside the MW potential illustrated above, with a leap-frog scheme. This is a modified Verlet integration method that is symplectic and suppresses numerical dissipation \citep{1967PhRv..159...98V, 2008gady.book.....B, 2003AcNum..12..399H}. The results from the orbital analysis are listed in Table ~\ref{tab:table2}.

\begin{figure}
	\includegraphics[width=\linewidth]{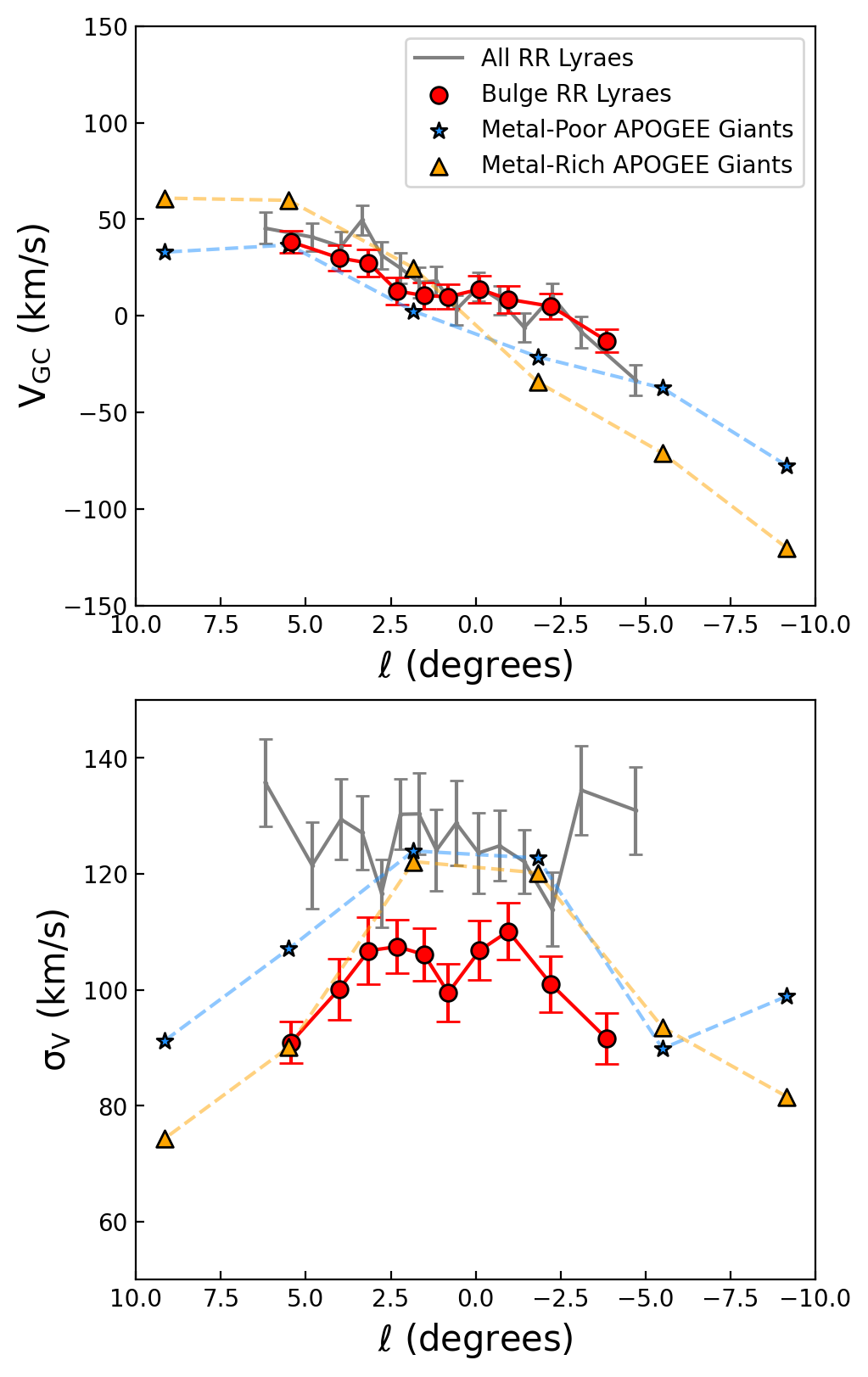}
    \caption{ Kinematics of the RRLs from this study. \textit{Top panel:} Rotation curve of the bulge-confined RRLs of our sample (red). The metal-poor (blue) and metal-rich (orange) rotation curves from \cite{Rojas2020} for giant stars are also presented for comparison. The rotation curve of all the RRLs of this study is included in grey. The bulge RRLs rotate more slowly than the metal-rich giants and are related to the metal-poor population. \textit{Bottom panel:} Velocity dispersion curves of the bulge RRLs (red) and the bulge giants. The velocity dispersion for all the RRLs is included in grey.}
    \label{fig:rotcurve}
\end{figure}
\section{Orbital classification and rotation curves}
\label{sec:rotation}
Using the orbital parameters obtained in the previous section, we separated the stars that are part of the bulge from those that are halo or disc interlopers. We used a new approach to do this by considering how long the stars stay within the bulge limits. We defined the bulge radius at 2.5 kpc. Based on the left panel of Fig. ~\ref{fig:orbselec}, we classified as a bulge star stars that during their orbital path remained inside the bulge for 80\% of the time at least. Conversely, we defined a halo star as a star that spends less than 30\% of the orbital period inside the bulge. Finally, we left as unclassified the stars that spend between 30 and 80\% of their orbits within 2.5 kpc. Stars in this regime could belong to the bulge, halo, and thick disc, and the uncertainties would increase if we included them in the other groups. Using these definitions, we found that 57\% (2385) of the sample are confined bulge RRL stars, 25\% (1069) are halo interlopers, and the rest are unclassified (742). The right plot of Fig. ~\ref{fig:orbselec} shows that in the frame of angular momentum versus energy, the bulge RRLs are well constrained. The halo and thick-disc stars are also more spread around these dimensions.

In order to derive the rotation curves of the bulge RRL stars, we divided the sample based on Galactic longitude into bins with a similar number of stars. Employing a bootstrap resampling technique, we computed a robust mean galactocentric radial velocity and velocity dispersion for each bin, along with their respective uncertainties. This method not only yields robust estimates of the mean and dispersion within each bin, but also provides reliable assessments of the associated errors.

The top panel of Fig. \ref{fig:rotcurve} shows the rotation curve of the confined bulge RRL stars (red). The rotation curve of all the RRLs of this study is plotted in grey. The bulge-confined RRLs show less rotation than the APOGEE metal-rich RC/RGB stars (orange) reported by \cite{Rojas2020} (at $\rm 2.5^\circ \leq |b| \leq 4.0^\circ $; see Fig. 15 in that work). This rotation is very similar but slightly slower than that of the metal-poor RC/RGB population (blue) of the same study. This result is the first evidence that relates the bulge RRLs with the bulge metal-poor population.  The bottom panel of Fig. \ref{fig:rotcurve} shows the velocity dispersion for the same populations. The velocity dispersion strikingly decreases significantly when only the bulge-confined RRLs are selected. It even has a similar profile (shape of its longitudinal variation), but is lower than that of the bulge RC/RGB populations overall. Most likely, this is due to the fact that we removed halo (and possibly thick-disc) stars from the sample, in agreement with the results of \cite{Lucey2021}.

\section{Three-dimensional spatial distribution}

\begin{figure}
	\includegraphics[width=\linewidth]{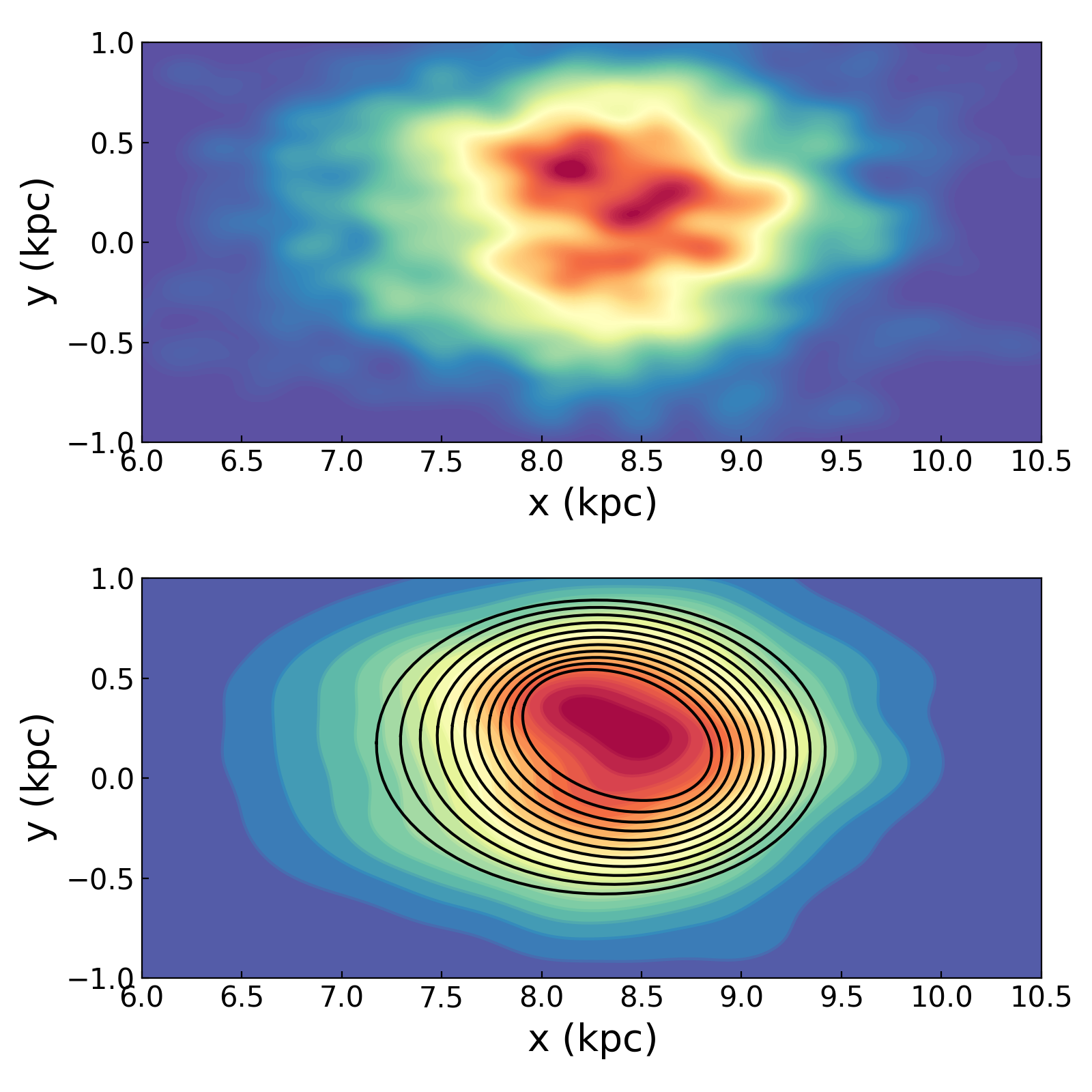}
    \caption{Position of the stars in the Galaxy. \textit{Top panel:} 2D KDE spatial distribution of the bulge RRL stars projected on the X-Y plane. \textit{Bottom panel}: Same spatial distribution with black ellipses adjusted to several contours.}
    \label{fig:3Ddist}
\end{figure}

We produced a 2D spatial distribution map using the Galactic coordinates and the distances derived in section \ref{sec:dist} for our RRLs sample by applying the \texttt{SkyCoord} routine from the Astropy package to obtain the Cartesian coordinates (X,Y,Z). Then, we applied a kernel density distribution (KDE) to the sample to highlight its shape. The top panel of Fig. ~\ref{fig:3Ddist} shows the face-on KDE spatial distribution of the bona fide bulge RRLs. Through the target selection, our sample is not uniform with longitude (see Fig. \ref{fig:location}). 
There are considerably more stars at positive than at negative $\ell$ , with a deficit close to $\ell=0^\circ$. In the bottom panel of Fig.~\ref{fig:3Ddist}, we show an ellipse fit to the KDE contours for completeness, but we consider the results unreliable due to the inhomogeneous coverage.

\begin{figure}
	\includegraphics[width=\linewidth]{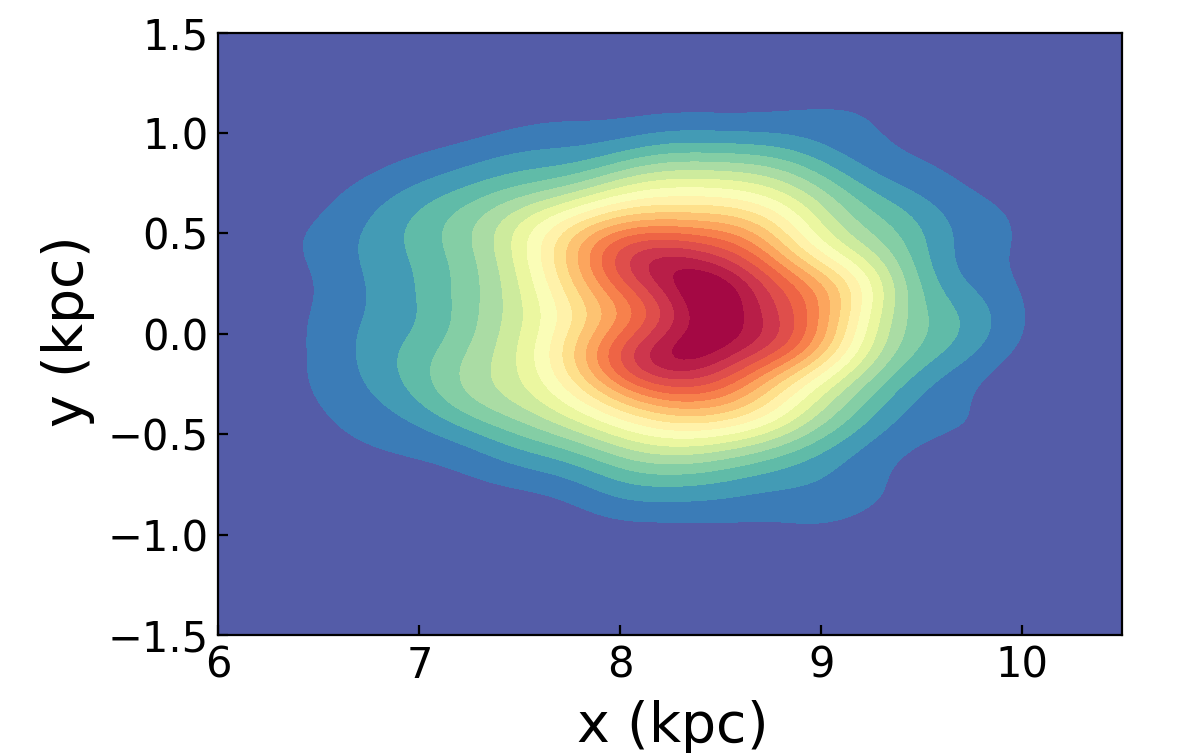}
    \caption{2D KDE spatial distribution of the bulge RRL stars projected on the X-Y plane with the symmetrization of the data by the stellar density.}
    \label{fig:3Ddist2}
\end{figure}

As an exercise, we tried to make the two sides of the distribution symmetrical by randomly removing stars at positive $\ell$, until we reached a similar density of stars on either side. Fig. ~\ref{fig:3Ddist2} shows the 2D KDE spatial distribution of this experiment. Our sample is still incomplete at $\ell=0^\circ$, but a more spheroidal distribution is observed. This result suggests that bulge RRL stars trace a rotating old spheroid. An extensive and more detailed study of the spatial distribution of RRL in the bulge region is presented in Zoccali et al. (2024). 

\section{Discussion and conclusion}

\begin{figure}
	\includegraphics[width=\linewidth]{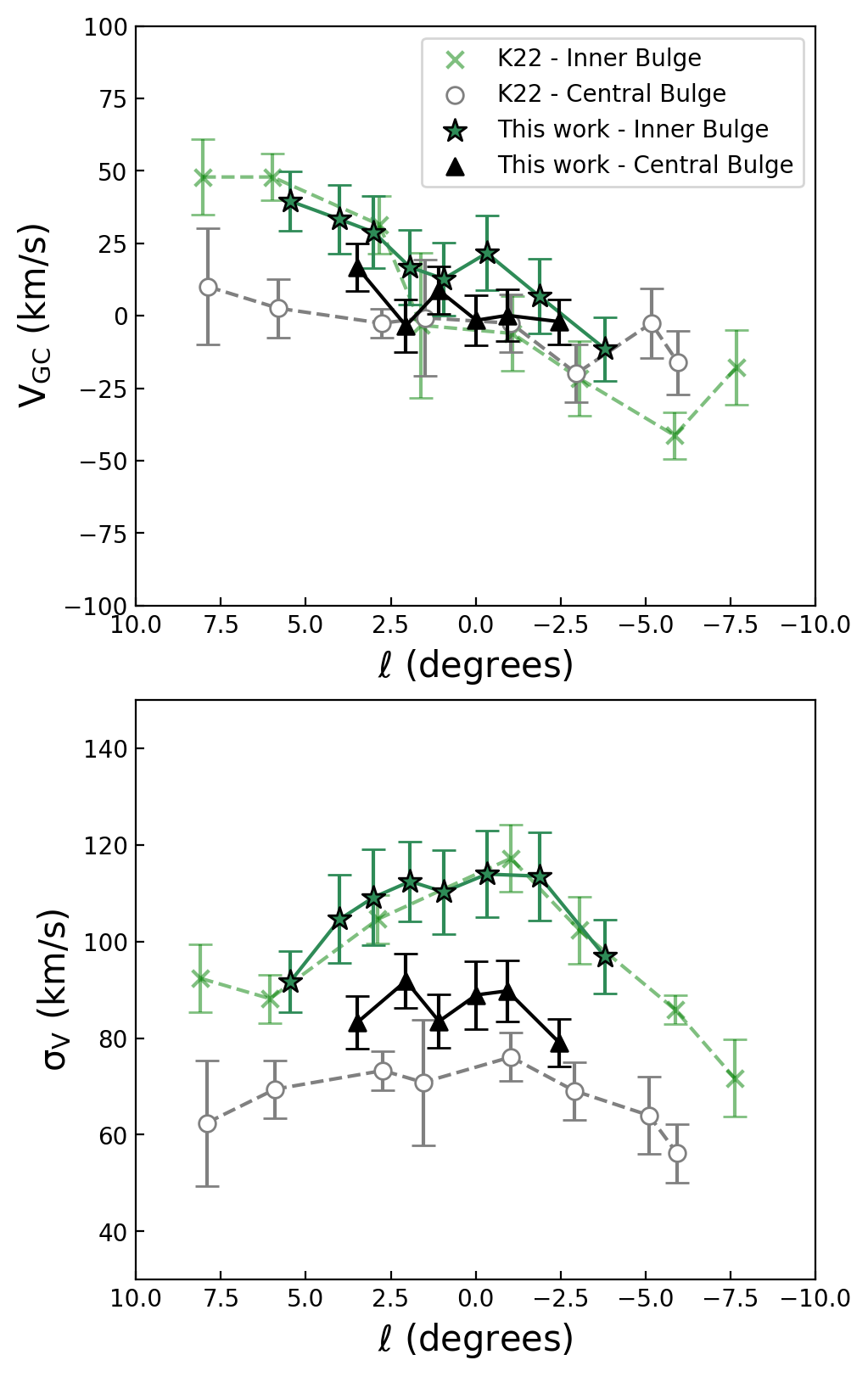}
    \caption{ Kinematics of the bulge RRLs compared with the results of K22. \textit{Top panel:} Rotation curve of the bulge-confined RRLs of our sample separated into those that are more centrally concentrated (black triangles) and those farther away (green stars). The rotation curves from K22 are also presented for comparison. Green shows the inner bulge and white the central bulge RRLs. We found a similar behaviour where inner bulge RRLs are rotating, and central ones do not show signs of rotation. \textit{Bottom panel:} Velocity dispersion curves for the distributions above. The velocity dispersion curve of the central bulge RRLs is clearly lower than that of the inner bulge ones, and it is flatter for both cases. On the other hand, the inner bulge ones show a velocity dispersion very similar to that of bulge giants (see Figure \ref{fig:rotcurve}).}
    \label{fig:1kpc}
\end{figure}

\begin{figure}
	\includegraphics[width=\linewidth]{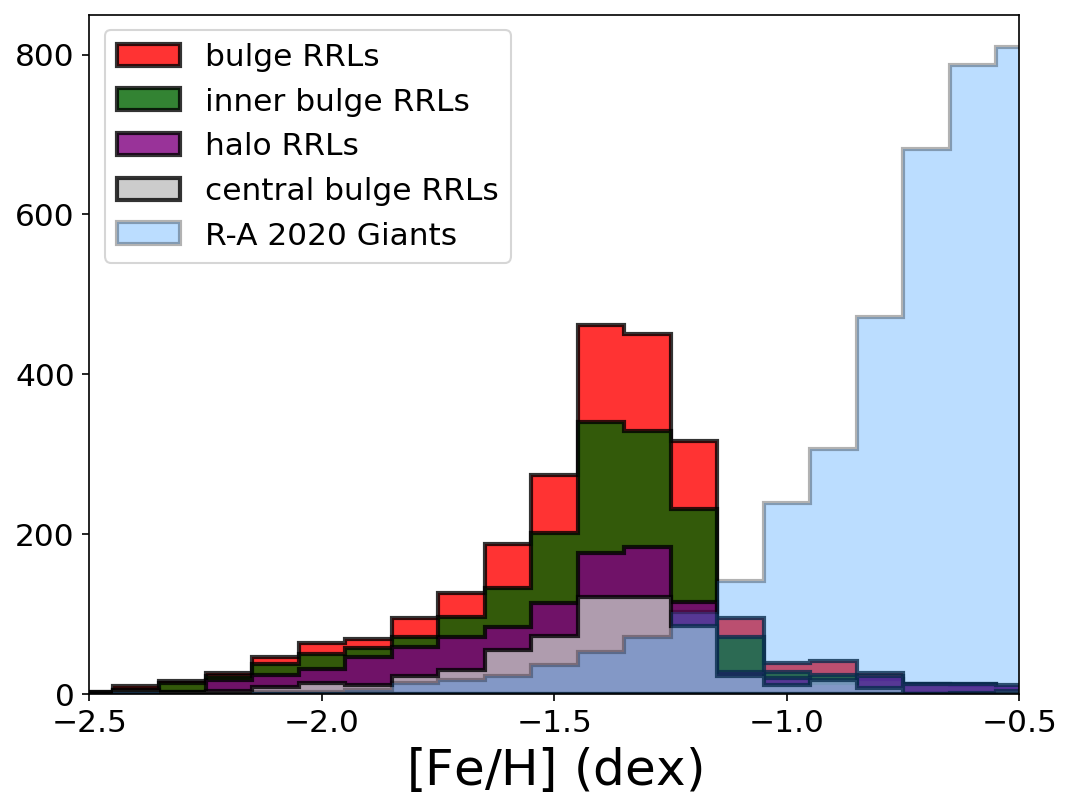}
    \caption{Metallicity distribution function of the bulge RRL stars (red), the halo and thick-disc stars (purple), the inner RRLs (green), and the central RRLs (grey). A peak at $-1.4$ dex is observed for the four distributions. The metallicities for APOGEE giants (sky blue) in the metal-poor regime are also shown.}
    \label{fig:metal}
\end{figure}

We analysed a sample of RRL confined to the Galactic bulge by combining RVs from APOGEE with light curves from OGLE and PMs from VVV. 
We derived systemic RV, distances, and transverse spatial velocities. We used an updated MW potential to obtain the orbital parameters for the individual stars in our sample. We used these parameters to isolate the stars that are confined to the bulge during most of their orbital paths. 

We produced the most complete catalogue of RRL stars in the Galactic bulge region to date, including RVs, PMs, mean magnitudes, metallicities, distances, and orbital parameters for more than 4200 stars. This catalogue is available online and can be used to search for old population structures, such as globular clusters or streams.

We found that 57\% of our initial sample are bulge-confined RRLs, 25\% are halo and disc interlopers, and the rest are unclassified. Previous studies found different contamination fractions in the bulge \citep[][K22]{Prudil2019}; the orbital analysis was also very different, however. Using a similar approach, \cite{Lucey2021} studied metal-poor bulge stars and found a contamination of 50\%, which is  very similar to our estimation. These results indicate that an orbital analysis is relevant for cleaning the sample.

We derived a rotation curve for the bulge RRLs and compared it with the metal-poor and metal-rich bulge populations as traced by RC/RGB stars. We found that the rotation velocity for RRL is similar to the velocity of metal-poor bulge RC/RGB stars and slower than that of metal-rich RC/RGB stars. The velocity dispersion decreases considerably when halo interlopers were removed. Its behaviour is also similar to that of other bulge populations. Although RRL stars are more metal-poor than RC stars on average, the relation between them is well explained by the fact that only stars belonging to the metal-poor tail of the metal-poor component burn helium in their core at a temperature that is high enough to cause them to cross the instability strip and pulsate. In other words, we find no evidence that RRL might trace a third bulge component.

Previous studies (K20 and K22) found that the more central RRLs do not rotate, while the others do rotate and follow the bar. In order to compare our results with theirs, we tried to separate two samples with similar criteria, but maintained our approach of a selection based on orbits rather than on the current position.
Specifically, we defined a central sample of stars that is confined within 1 kpc from the Galactic centre for at least 80$\%$ of their orbital period and an inner-bulge sample of stars that is confined within 2.5 kpc for the same percentage of their orbital time. As a result, 1766 (74\%) out of 2385 bulge RRLs are in the inner bulge, and 619 (26\%) are in the central region. For a consistent comparison, we tried to apply our approach to the K22 data by re-calculating orbits with our code, using their distances and velocities. This resulted in 936 RRL in the inner bulge and 113 in the central region. The latter number is too small to allow us a robust comparison in longitude bins. We show this comparison in Appendix A, but here, we prefer to compare our results directly with the K20/K22 selection, even through the adopted criteria are different. 

The upper panel of Fig. ~\ref{fig:1kpc} shows the rotation curves obtained with our sample, adopting the cut at 1 kpc and comparing with the results by K22 (white circles: central RRLs; light green crosses: inner RRLs). The rotation curve with black triangles is our result for the central RRLs, while the dark green stars show the rotation curve for the rest of the stars from the bulge. The results are very similar to those of K22. The central RRLs in our sample show no net rotation, while the inner RRLs clearly rotate. It is unclear why the central RRLs do not rotate. We need more data from, for instance, RRLs closer to the Galactic plane to increase the statistics and answer this question. 

The velocity dispersion curves are presented in the bottom panel of Fig. \ref{fig:1kpc}. For the inner RRLs, the results are identical to those of K22, showing a behaviour comparable to that of the APOGEE giants (Fig. \ref{fig:rotcurve}, bottom panel). On the other hand, the central RRLs present a lower velocity dispersion than those of the inner bulge, but our result presents a higher velocity dispersion compared with the result of K22, although most of the bins are between the error bars. The velocity dispersion for K22 may not be robust enough since they had fewer stars per bin. Moreover, our result is consistent with previous studies \citep{Arentsen2020,Arentsen2023} for very metal-poor stars. 

Fig. \ref{fig:metal} shows the metallicity distribution of the RRLs in our sample. The repeated division of the RRLs into a central (grey) and inner (green) sample shows no metallicity difference. Furthermore, bulge (red) and halo and thick-disc RRLs (purple) show similar metallicity distributions. The four distributions span the complete metallicity range, with a peak always close to $-$1.4 dex.
Within the errors of the current photometric metallicities, we find no differences in the metallicity distribution of the different spatial/kinematical samples. We therefore conclude that photometric metallicity is not a suitable parameter for separating RRL populations in the Galactic bulge region. 

Concerning the spatial distribution, while our sample seems to support a spheroidal distribution as opposed to a bar, we emphasize that our data do not allow us to reach a robust conclusion because the longitude sampling of the target selection was non-uniform. For a further discussion of the spatial distribution of RRL stars in the bulge, we refer to Zoccali et al. (2024).

Regarding the apparent rotation of the bulge component traced by the RRLs of our sample, our work adds to the wealth of evidence for rotating bulges in spiral galaxies \citep[i.e.][]{Kormendy1982, Cappellari2007, Fabricius2012}. The origin of a rotating structure like this is still debated, as there is evidence from Galactic evolution models of both rotating classical (accreted) bulges \citep[][i.e. model Aq-D-5]{Tissera2018} and bulges created in situ with rotation since their birth \citep[][i.e.]{DiMatteo2016,Debattista2017}. Since a detailed study of the origin of the old spheroid is beyond the scope of this paper, we limit our investigation in this direction to pointing out that the amount of rotation we observe in the RRL sample can be explained by the angular momentum transfer from the bar through secular evolution (i.e. \citealt{Saha2012}, and especially the models RCG100 and RCG101 from \citealt{Saha2016}).


\begin{acknowledgements}

J.O.C. acknowledges Andrea Kunder for the extensive conversations about the topics included in this manuscript and also for her availability to provide her data and advice for the future.

Based on observations taken within the ESO VISTA Public Survey VVV, Program ID 179.B-2002, made public at the ESO Archive and through the Cambridge Astronomical Survey Unit (CASU).

J.O.C. acknowledges support from the National Agency for Research and Development (ANID) Doctorado Nacional grant 2021-21210865, and by ESO grant SSDF21/24. 

This work is funded by ANID, Millenium Science Initiative, ICN12\_009 awarded to the Millennium Institute of Astrophysics  (M.A.S.), by the ANID BASAL Center for Astrophysics and Associated Technologies (CATA) through grant FB210003, and by  FONDECYT Regular grant No. 1230731. A. R. A. acknowledges support from DICYT through grant 062319RA. C. Q. Z. acknowledges support from the National Agency for Research and Development (ANID), Scholarship Program Doctorado Nacional 2021 – 21211884, ANID. E. V. acknowledges the Excellence Cluster ORIGINS Funded by the Deutsche Forschungsgemeinschaft (DFG, German Research Foundation) under Germany’s Excellence Strategy – EXC-2094-390783311. A. V. N. acknowledges support from the National Agency for Research and Development (ANID), Scholarship Program Doctorado Nacional 2020 – 21201226, ANID.

We gratefully acknowledge the use of data from the OGLE-IV catalogue.
The OGLE project has received funding from the National Science Centre, Poland, grant MAESTRO 2014/14/A/ST9/00121 to AU.

We also acknowledge the use of data from the VVV/VVVx
ESO Public Survey program ID 179.B-2002/198.B-2004 taken
with the VISTA telescope and data products from the Cambridge
Astronomical Survey Unit (CASU). The VVV Survey data are made
public at the ESO Archive.

Funding for the Sloan Digital Sky Survey IV has been provided
by the Alfred P. Sloan Foundation, the U.S. Department of Energy
Office of Science, and the Participating Institutions. SDSS-IV
acknowledges support and resources from the Center for High-
Performance Computing at the University of Utah. The SDSS
website is www.sdss.org.

SDSS-IV is managed by the Astrophysical Research Consortium for the Participating Institutions of the SDSS Collaboration
including the Brazilian Participation Group, the Carnegie Institution
for Science, Carnegie Mellon University, the Chilean Participation
Group, the French Participation Group, Harvard-Smithsonian Center
for Astrophysics, Instituto de Astrofisica de Canarias, The Johns
Hopkins University, Kavli Institute for the Physics and Mathematics
of the Universe (IPMU) University of Tokyo, Lawrence Berkeley National Laboratory, Leibniz Institut für Astrophysik Potsdam
(AIP), Max-Planck-Institut für Astronomie (MPIA Heidelberg),
Max-Planck-Institut für Astrophysik (MPA Garching), Max-Planck Institut für Extraterrestrische Physik (MPE), National Astronomical
Observatory of China, New Mexico State University, New York
University, University of Notre Dame, Observatário Nacional/MCTI, 
The Ohio State University, Pennsylvania State University, Shanghai
Astronomical Observatory, United Kingdom Participation Group,
Universidad Nacional Autónoma de México, University of Arizona,
University of Colorado Boulder, University of Oxford, University of
Portsmouth, University of Utah, University of Virginia, University
of Washington, University of Wisconsin, Vanderbilt University, and
Yale University.

It also made use of NASA’s Astrophysics Data System and of the VizieR catalogue access tool, CDS, Strasbourg, France \citep{simbad}.  The original description of the VizieR service was published in \citep{vizier}. 
Finally, we acknowledge use of the following publicly available softwares: lcfit: A python package for the regression of periodic time series \citep{Dekany2019}, rr\_feh \citep{Dekany2022}, PyFiNeR: Fitting Near-infrared RRL light curves \citep{Hajdu2018}, TOPCAT \citep{topcat}, pandas \citep{pandas}, IPython \citep{ipython}, numpy \citep{numpy}, matplotlib \citep{matplotlib}, Astropy, a community developed core Python package for Astronomy \citep{astropy1,astropy2} and Aladin sky atlas \citep{aladin1, aladin2}. 

\end{acknowledgements}



\bibliographystyle{aa}
\bibliography{mybiblio} 

\begin{appendix}

\section{Comparison with the BRAVA-RRL2 dataset}

\begin{SCfigure*}[0.5][htbp!]
\caption{Kinematics of the bulge RRLs compared with the data of K22 using our approach. \textit{Top panels:} Rotation curve of the bulge-confined RRLs of our sample separated by those farther away (dashed red line) and those more centrally concentrated (dashed grey line). The rotation curves from K22 using our orbital selection are also presented where with green crosses are the inner bulge and white circles for the central bulge RRLs. \textit{Bottom panels:} The velocity dispersion curves using the same symbols. The velocity dispersion of the central bulge RRLs from K22 have larger errors because there are fewer stars per bin.}
\includegraphics[width=0.6\textwidth]{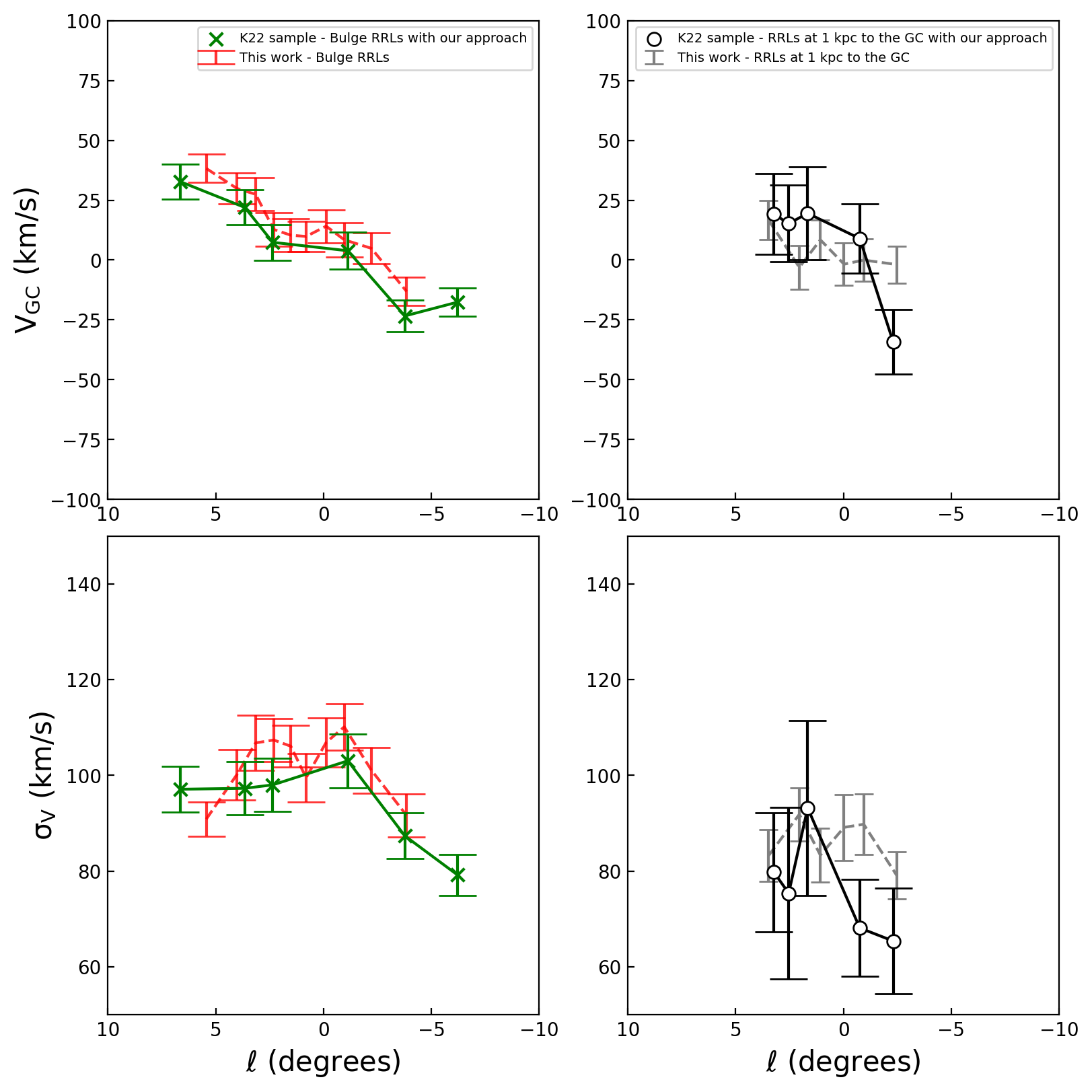}
\label{fig:appendix2}
\end{SCfigure*}
 We hereby present a different comparison with K22, for which we re-derived the orbital parameters of their stars using the position and velocities from their tables. We then defined an inner and a central sample with the same criteria as we applied to our sample. The comparison is conceptually more consistent than that presented in Fig.~\ref{fig:1kpc}, but the statistics for the central RRLs in their sample are very low, as only 113 stars satisfied the inner criterion.
Fig. \ref{fig:appendix2} shows the resulting rotation curves. The rotation of the inner RRL in K22 (green crosses) is very similar to that for our inner RRL (red), while the central RRL in K22 (black) show almost no rotation, although with very large error bars. The same is seen in the velocity dispersion (right panels), where the curves for our sample and K22 are very similar, although the latter has significantly lower statistics, especially in the central sample.
\FloatBarrier

\FloatBarrier

\section{Testing the relevance of the errors on the orbital results}

As shown in the right panel of \ref{fig:orbselec}, our selection based on the full orbital modelling is similar to a selection based on the present-day total orbital energy, $E_{tot}$, but it is conceptually more accurate. This is because our selection is purely kinematical, so that a given observed position and velocity will unambiguously determine whether a star is included in or excluded from the bulge sample. A selection in energy has basically the same effect, except in a transition region (the area where the red and grey sample overlap in the right panel of \ref{fig:orbselec}), where only the full orbit integration can distinguish between genuine bulge stars and interlopers. 

In order to quickly test the impact of the errors in the systemic velocity, distance, and proper motions, we produced copies of our original sample, iterating over the errors and then repeated the rotation and velocity dispersion curves, selecting only stars with $\rm E_{tot} \leq -1.9\times10^5 \ km^2 s^{-2}$. This cut included $>90\%$ of our originally selected bulge sample. We iterated over the errors by producing copies of each star, for which the variables ($V_{GC}$, distance, $\mu_l$, and $\mu_b$) were extracted from Gaussians with a centre value equivalent to the observed value and a standard deviation equal to the measured uncertainties. We repeated this process 1000 times for all stars in our sample, thus producing 1000 copies of our original $\sim$4000 stars, and then selected for each new sample a bulge subsample based on the total orbital energy cut mentioned above.

\begin{figure}
	\includegraphics[width=\linewidth]{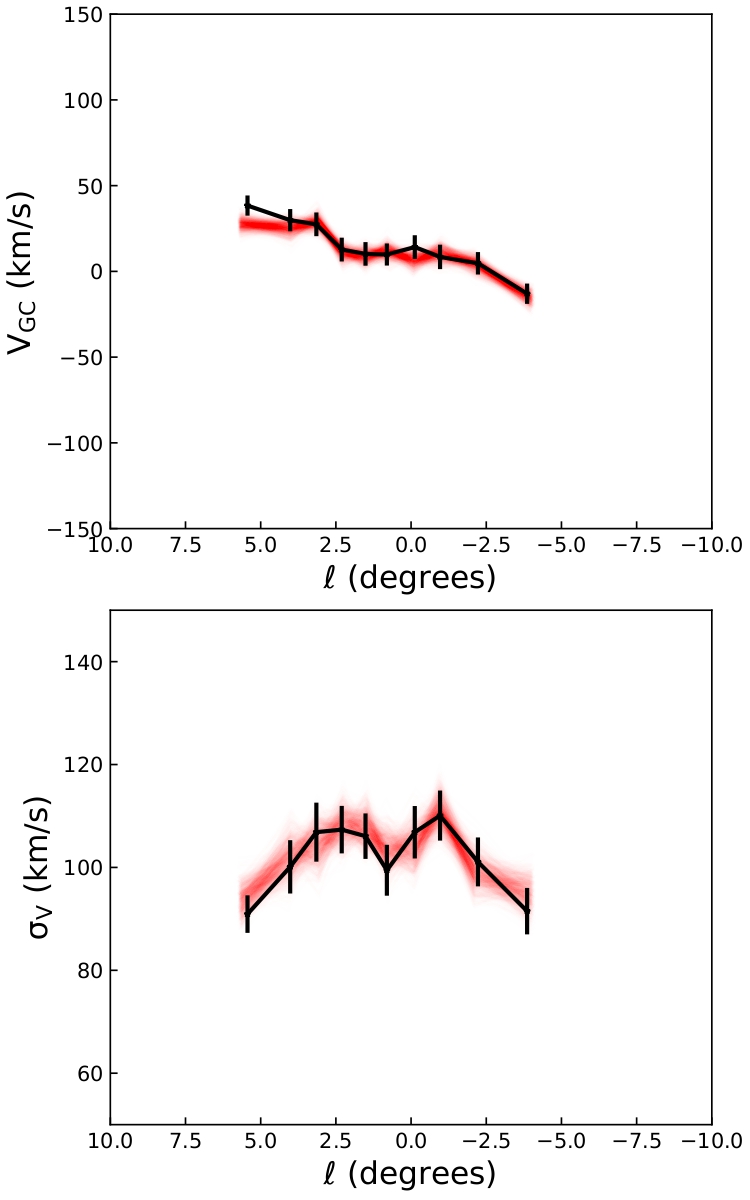}
    \caption{Influence of the errors in the rotation curve and velocity dispersion versus longitude. \textit{Top panel:} Rotation curve of the bulge-confined RRLs of our sample (black line) and in faint red the 1000 copies of the rotation curve based on random errors in systemic RV, PMs, and distance. \textit{Bottom panel:} Velocity dispersion curves using the same symbols.}
    \label{fig:errors}
\end{figure}

The final result of this procedure is shown in \ref{fig:errors}, which is a copy of \ref{fig:rotcurve}, where we have plotted in semi-transparent red the curve for each copy sample and a black curve for our original results. In the areas that appear redder, the results of most of the copies overlap, and thus, these areas represent the most probable result even when the errors are taken into account. \ref{fig:errors} shows that the agreement between the redder areas and the black lines is remarkable (it also shows the goodness of bootstrapping that was used originally in mimicking the errors). The slight tension between the red areas and the black line for the bin at higher galactocentric distance (at $\ell>5^\circ$) arises because this corresponds to the transition region of the total energy distribution mentioned above. It is crucial to emphasize that the random sampling over the errors does not always give the same number of stars in the bulge sample, nor does it always give the exact same stars, especially for copies of stars whose energies lie close to the threshold used for the selection. Nonetheless, the agreement of this technique with our results is reassuring, considering that this iteration over the errors represents an overestimation of the impact of errors, given that the energy selection is worse at selecting bulge stars than the selection that exploits the full orbital analysis that we employed.

\end{appendix}
\label{lastpage}
\end{document}